\documentstyle[12pt]{article}
\begin{document}
\thispagestyle{empty}

\begin{center}
\LARGE \tt \bf{Spherically symmetric space-time defect solution of Einstein field equations.}
\end{center}

\vspace{2.5cm}

\begin{center} {\large L.C. Garcia de Andrade\footnote{Departamento de
F\'{\i}sica Te\'{o}rica - Instituto de F\'{\i}sica - UERJ

Rua S\~{a}o Fco. Xavier 524, Rio de Janeiro, RJ

Maracan\~{a}, CEP:20550-003 , Brasil.

E-Mail.: GARCIA@SYMBCOMP.UERJ.BR}}
\end{center}

\vspace{2.0cm}

\begin{abstract}
A new class of spacetime defect solutions of Einstein Field equations of Edelen's direct Poincar\'{e} Gauge Field theory without biaxial symmetry is presented. The interior solution describes a core of defects where curvature vanishes and Cartan torsion is nonvanishing.
Outside the core (in vacuum) the solution represents a spacetime with vanishing curvature and torsion describing a nontrivial topological defect solution of Einstein equations of gravity. Our solution corresponds to a very weak strenght of Tachyons can be found far away from the core defect.
\end{abstract}

\newpage

\section{Introduction}

Recently D.G.B.Edelen \cite{1} has presented a class of spacetime defect solutions of Einstein field equations of general relativity with biaxial symmetry where defect core functions are homogeneous of degree - 2 yielding a vanishing torsion. Riemann curvature is assumed to vanish and therefore the solution is a solution of Einstein field equation in vacuum $ R_{ik}=0 $ , although being topological nontrivial.
In this paper I shall be concerned with the investigation of a new class of defect solutions of Einstein field equations without biaxial symmetry.
Inside the defect core curvature vanishes and only Cartan's torsion is nonzero.
More recently several \cite{2,3,4,5,6} spacetime defect metrics have been given as solutions of Einstein-Cartan field equations of general relativity with spin and torsion.
One of these solutions describes defects in Weitzenb\"{o}ck spacetime \cite{5}. We also compute the geodesics for the vacuum part of the solution and show that there is a tachyonic sector as in previously Edelen \cite{1} paper.

\section{Gauge Differential Geometry.}

In 1986 and 1989 Edelen described the following results.
The Minkowski spacetime $ M_{4} $ with global coordinates $ \{x,y,z,t \} $ is the base of a $ L_{4} $ Riemann-Cartan Spacetime which is generated from the action of a Poincar\'{e} group on $ M_{4} $. The Riemann-Cartan manifold is , in general, endowed with both curvature and torsion. The translation group T(4) yields the compensating one-forms $ {\phi}^{i}={{\phi}^{i}}_{j}(x^{k})dx^{j} \hspace{0.5cm}(1 \le i \le 4) $ and local axial of the six-parameter L(6) local Lorentz group and ten-parameters Poincar\'{e} group P(10)$ \subset $GL(5,R)are given by

\begin{eqnarray}
{W}^{\alpha}={W^{\alpha}}_{i}(x^{k}) dx^{i} \hspace{1.5cm} (1 \le {\alpha} \le 6) \\
{B}^{i} = {{B}^{i}}_{j}(x^{k})dx^{j}=( {{\delta}^{i}}_{j}+ {{W}^{\alpha}}_{j} {{l}^{i}}_{k \alpha} {x}^{k}+{{\phi}^{i}}_{j} )d{x}^{j}
\end{eqnarray}
respectively. The distortion 1-forms $ \{ B^{i} \vert 1 \le i \le 4 \}$ are the basis of a vector space $ \wedge ^{1} $ of forms on $ L_{4} $.

The distorted Riemann-Cartan spacetime $ L_{4} $ obtained from $ M_{4} $ by minimal substitution yields the line element

\begin{equation}
ds^{2}=g_{ij} dx^{i} \otimes  dx^{j}
\label{3}
\end{equation}
where $ g_{ij}={{B}^r}_{i}h_{rs}{{B}^s}_{j}  ,   g=det(g_{ij})=-B^{2} $ and $ ds^{2}=h_{ij}dx^{i} \otimes  dx^{j} $ is the $ M_{4} $ line element.

The spacetime $ L_{4} $ has both curvature and torsion in general. The Cartan torsion 2-forms $ \{ {\sum}^{i}| 1 \le i \le 4 \} $ are given by 

\begin{equation}
{\sum}^{i}=dB^{i}+W^{\alpha} {{l^{i}}_{j \alpha}} \wedge B^{j}
\label{4}
\end{equation}

Where the holonomic torsion 2-forms $ S^{k}= \frac{1}{2}({{\Gamma}^{k}}_{ij}-{{\Gamma}^{k}}_{ji}) dx^{i} \wedge dx^{j} $ are determined in terms of the $ \sum^{i} $ by  $ S_{k}={{b}^{k}}_{r} {\sum}^{r} $ where $ b_{i}  \rfloor   B^{j}={{\delta}^{j}}_{i} $ ,\linebreak $ b_{i}={{b}^{j}}_{i}(x^{k}) {\partial}_{j} $ being the frames of $ B^{j}$ .
In general the torsion forms are given by (the coframes)

\begin{equation}
{\sum}^{i}={\theta}^{\alpha} {l^{i}}_{j \alpha} {\chi}^{j} + d{\phi}^{i} + W^{\alpha} {l^{i}}_{j \alpha} \wedge {\phi}^{j}
\label{5}
\end{equation}
where $ {\theta}^{\alpha}= \frac{1}{2}{{\theta}^{\alpha}}_{rs} dx^{r} {\wedge} dx^{s} $ and the Riemann curvature is given by $ {R^{i}}_{rsj} = {{\theta}^{\alpha}}_{rs} {L^{i}}_{j \alpha} $.
In this paper we shall  be concerned with dislocations where curvature vanishes and only torsion survives.
Thus $ {\theta}^{\alpha} = 0 $ , $ {R^{i}}_{rsj} = 0 $. 
Defining $ W^{\alpha} \equiv 0 $ the dislocation density and current (Cartan torsion) reduces to $ {\sum}^{i} = d{\phi}^{i} $ and the distortion 1-forms have the form $ B^{i}=dx^{i} + {\phi}^{i} $.
In general in crystalline solids the procedure consists in giving the dislocation density 2-forms and then to calculate the response of the solid.
Here we shall consider a dislocation density like

\begin{equation}
{\sum}^{i}= A^{i}(R,t) dR{\wedge}dt
\label{6}
\end{equation}
From the expression $ {\sum}^{i}=d{\phi}^{i} $ , $ d{\sum}^{i}= 0 $.
On integration of the system yields

\begin{equation}
{\phi}^{i}=a^{i}(R,t)(Rdt-tdR)
\label{7}
\end{equation}
the essential difference between these functions here and Edelen's functions in \cite{1} is that the functions here are not biaxial functions \cite{7}. The functions (\ref{7}) are indeed homogeneous of degree -2 outside the core of defects since Cartan torsion is

\begin{equation}
{\sum}^{i}=d{\phi}^{i}=\{ \frac{{\partial}a^{i}}{{\partial}R}R + \frac{{\partial}a^{i}}{{\partial}t}t + 2 a^{i} \} dR {\wedge} dt
\label{8}
\end{equation}
and therefore the region $ R > R_{0} $ (here $ R=\sqrt{x^{2}+y^{2}+z^{2}} $ is a homogeneous function of degree 1) if $ a^{i} $ are homogeneous of degree -2, $ {\sum}^{i}=0 $ from (\ref{8}) and curvature and torsion vanish. Despite of this situation the solution of Einstein field equation in vacuum $ (R_{ik}=0) $ is topologically nontrivial like the ones \cite{9} obtained earlier by Marder in the context of general relativity and by Tod \cite{8} and Letelier \cite{3,4} in the context of Einstein-Cartan theory of gravity.

\section{Spherically Symmetric Dislocation in Spacetime.}

To obtain the metric form of the above solution and to investigate the geodesics one needs to compute the frame $ \{ b^{i} \} $ basis which yields (here we have consider the approximation where $ O(f^{2}) \to 0 $, where the strenght of dislocation is very weak)

\begin{eqnarray}
B^{1} & = & -(1-fR)dR-f \frac{R}{t} dt \\
B^{2} & = & d{\theta}+a \hspace{0.5cm},\hspace{0.5cm} B^{3}=d{\varphi} \\
B^{4} & = & (1+fR)dt-ftdR
\end{eqnarray}
and

\begin{eqnarray}
b_{1} & = & -(1+fR){\partial}_{R}+ft {\partial}_{t} \\
b_{2} & = & {\partial}_{\theta} \hspace{0.75cm},\hspace{0.75cm} b_{3}={\partial}_{\varphi} \\
b_{4} & = & - \frac{fR}{t} {\partial}_{R}+(1-fR){\partial}_{t}
\end{eqnarray}
where we have used the result $ a^{1} \equiv - f $ and $ Ra^{1}=ta^{4}$. From the frame equations we obtains the line element

\begin{equation}
ds^{2}=+(1+fR)^{2}dt^{2} - (1-fR)^{2} dR^{2} - R^{2}(d{\theta}^{2}+ {\sin}^{2}{\theta} d{\varphi}^{2})-2ft(1+\frac{1}{2}Rf+ft)dRdt
\label{15}
\end{equation}
Notice that for $ f \equiv - \frac{GM}{R^{2}} $ and $ O(f^{2}) \to 0 $ ,(\ref{15}) reduces to Schwarzschild metric. Nevertheless this is not a solution to our problem since Shwarzschild solution although is a solution of Einstein field equation in vacuum has a nonvanishing Riemann curvature. Sphericalbubbles of this type have been consider by Letelier and Wang \cite{10}. In our case one must define $ f \equiv \frac{K}{R^{2}} $ where K is a dislocation strength being zero outside the core defect. Therefore $ f $ must vanish outside the core defect and the metric (\ref{15}) will be flat outside the core defect.
This metric is nonsingular since $  det(g_{ij}) \ne 0 $ as can be easily checked. Metric (\ref{15}) fits into the general spherically symmetric form

\begin{equation}
ds^{2}=A(r,t)dt^{2} - B(r,t) dr^{2} - r^{2}(d{\theta}^{2}+{\sin}^{2}{\theta} d{\phi}^{2})- 2F(r,t)drdt
\label{16}
\end{equation}
as it is known from Riemannian geometry this metric can be reduced to a static metric by a change of coordinates. Therefore we are left with a spherically symmetric solution of the Einstein field equation. To compute the geodesic equations

\begin{equation}
{\dot{v}}^{i}=0 \hspace{1.5cm} {\dot{u}}^{i}={b^{i}}_{j} v^{j}
\label{17}
\end{equation}

Let us define the vector $ v^{i} $ from the first equation in (\ref{17}) as

\begin{equation}
v^{i}=(0,k,0,1)
\label{18}
\end{equation}
where k is a constant. From the velocity expression $ V^{i}={{b}^{i}}_{j}{v}^{j} $ and (\ref{18}) one obtains

\begin{equation}
V^{i}=(0,k,0,(1-fR))
\label{19}
\end{equation}

To obtain the geodesic equations we substitute (\ref{19}) into the equations

\begin{equation}
v^{i} ={{B}^{i}}_{k}V^{k}
\label{20}
\end{equation}
and use these into the second eqn. in (\ref{17}) obtaing

\begin{eqnarray}
\frac{dR}{d{\tau}} & = & ft \\
\frac{dt}{d{\tau}} & = & (1-fR) \\
\frac{d{\theta}}{d{\tau}} & = & k \hspace{.35cm},\hspace{.35cm} \frac{d{\varphi}}{d{\tau}}=0
\end{eqnarray}

Since our aim is to show the existence of tachyons even in the linear approximation of the strenght of dislocation $ K(O(K^{2})>0) $ we made the simplest choice for the torsion function $ f $ or $ f \equiv \frac{K}{R^{2}} $. From metric (\ref{15}) one may notice that the solution $ f \equiv K $ would led to a non-Minkovski metric as $ R $  goes to infinite. With this choice the geodesic equations can be rewritten as 

\begin{eqnarray}
\frac{d^{2}R}{d{\tau}^{2}} & = & K(\frac{1}{R}-\frac{K{\tau}}{R^{2}}) \hspace{.5cm} (R(0)=R_{0}) \\
\frac{d^{2}t}{d{\tau}^{2}} & = & \frac{K^{2}t}{R^{4}} \approx 0 \\
\frac{d^{2}{\theta}}{d{\tau}^{2}} & = & 0 \hspace{.35cm},\hspace{.35cm} \frac{d^{2}{\varphi}}{d{\tau}^{2}}=0
\end{eqnarray}

Which integrate to

\begin{eqnarray}
R({\tau}) & = & \frac{K{\tau}^{2}}{2}+{R_{0}} \\
t({\tau}) & = & {\tau} \hspace{0.35cm},\hspace{0.35cm} {\theta}({\tau})=K \hspace{0.35cm},\hspace{0.35cm} {\varphi}({\tau})=const.
\end{eqnarray}

\begin{equation}
V^{2} \equiv V^{i}g_{ij}V^{j}=v^{i}h_{ij}v^{j}=(1-k^{2}R^{2})>0
\label{29}
\end{equation}
thus this observer is a proper test particle for the spacetime $ L_{4}$.

Notice that an observer in the asymptotic Minkovski space at infinity would obtain $ V^{i}h_{ij}V^{j} \cong [1-2KR_{0} (1-\frac{k^{2}R_{0}}{2K})] $ around $ \tau = 0 $. The spatial part of thus velocity would be 

\begin{equation}
V^{2}=2KR_{0}(1- \frac{k^{2}R_{0}}{2K})
\label{30}
\end{equation}

From formula (\ref{30}) it is easy to note that the region $ R_{0}= \frac{2K}{k^{2}} $ is forbidden for tachyons. Around this spherical surface tachyons are not forbidden. Thus there is a possibility to find tachyons around weak spherical spacetime defect cores. Letelier and Wang \cite{10,11} have investigated spherically symmetric spacetime defects without torsion where Riemann-Christoffel is novanishing only at surface defects. In Letelier-Wang's \cite{10} spherical defects no tachyons appear.
Another defect solutio given by $ f = \frac{K}{R^{2}} $ would led us to tachyons around the core defect. Other defect geometries and their relation to tachyons can appear elsewhere.

\section*{Acknowledgement}
I am very much indebt to Prof. D.G.B. Edelen for Rindly suggesting this problem to me and to CNPq. and UERJ for financial support. Thanks are also due to Prof. W.A.Rodrigues Jr. for discussions on tachyons.

\end{document}